\documentclass[twocolumn, showpacs, preprintnumbers,amsmath,amssymb]{revtex4}

\usepackage{amsfonts}
\usepackage{dsfont}
\usepackage{amsxtra}
\usepackage{hyperref}

\usepackage{graphicx,epsfig}
\usepackage{dcolumn}
\usepackage{bm}

\newcounter{theappend}

\newcommand{\be}{\begin{equation}}
\newcommand{\ee}{\end{equation}}
\newcommand{\bea}{\begin{eqnarray}}
\newcommand{\eea}{\end{eqnarray}}
\newcommand{\beaa}{\begin{eqnarray*}}
\newcommand{\eeaa}{\end{eqnarray*}}

\newcommand{\ba}{\begin{array}}
\newcommand{\ea}{\end{array}}
\newcommand{\bi}{\begin{itemize}}
\newcommand{\ei}{\end{itemize}}
\newcommand{\ben}{\begin{enumerate}}
\newcommand{\een}{\end{enumerate}}

\newcommand{\bra}{\langle}
\newcommand{\ket}{\rangle}
\newcommand{\ra}{\rightarrow}

\newcommand{\lb}{\label}

\newcommand{\al}{\alpha}

\newcommand{\p}{\partial}
\newcommand{\dl}{\delta}

\newcommand{\ld}{\lambda}

\newcommand{\sm}{\sigma}

\newcommand{\vx}{{\bf x}}

\newcommand{\mdm}{M_{\rm DM}}
\newcommand{\rhodm}{\rho_{\rm DM}}
\newcommand{\VLII}{{\it Via Lactea~II}}

\begin{document}

\title{ATIC, PAMELA, HESS, Fermi \\
and nearby Dark Matter subhalos}
\author{Michael Kuhlen}
 \email{mqk@ias.edu}
 \affiliation{Institute for Advanced Study, Einstein Drive, Princeton, NJ 08540}
\author{Dmitry Malyshev}
 \email{dm137@nyu.edu}
 \altaffiliation{On leave of absence from ITEP, Moscow, Russia, B. Cheremushkinskaya 25}
\affiliation{ CCPP, 4 Washington Place, Meyer Hall of Physics, NYU, New York, NY 10003}

\date{\today}

\begin{abstract}
\noindent 
We study the local flux of electrons and positrons from annihilating Dark Matter (DM), and investigate how its spectrum depends on the choice of DM model and inhomogeneities in the DM distribution. Below a cutoff energy,
the flux is expected to have a universal power-law form with an index $n \approx -2$.
The cutoff energy and the behavior of the flux near the cutoff 
is model dependent.
The dependence on the DM host halo profile may be significant at energies $E < 100$ GeV
and leads to softening of the flux, $n < -2$.
There may be additional features at high energies due to the presence of local clumps of DM, especially for models in which the Sommerfeld effect boosts subhalo luminosities.
In general, the flux from a nearby clump gives rise to a harder spectrum of electrons
and positrons, with an index $n > -2$.
Using the {\VLII}
simulation, we estimate the probability of such subhalo effects
in a generic Sommerfeld-enhanced model to be at least 4\%, and possibly as high as 15\% if subhalos below the simulation's resolution limit are accounted for.
We discuss the consequences of these results 
for the interpretation of the ATIC, PAMELA, HESS, and Fermi data, 
as well as for future experiments.

\end{abstract}


\maketitle

\newpage

\tableofcontents


\section{Introduction}

One of the corner stones of the standard big bang cosmology is the
presence of dark matter (DM).  Although DM comprises at least 80\% of
all matter in the Universe, surprisingly little is known about its
nature.  A recent rise of interest in the DM problem has been
triggered by observations of anomalies in several high energy
cosmic ray experiments.  The PAMELA Collaboration
\cite{Adriani:2008zr} reported an increasing positron fraction in the
energy range 10 - 100 GeV, while the ATIC \cite{:2008zzr}, PPB-BETS
\cite{Torii:2008xu}, and HESS \cite{Collaboration:2008aaa} experiments
suggest a bump in the total $e^+e^-$ flux between 100 GeV and 1 TeV.
Recent Fermi/LAT \cite{Collaboration:2009zk} 
and HESS \cite{Aharonian:2009ah} data show a smooth
spectrum which is harder than the expected background at energies
$E > $ 100 GeV and has a break around 1 TeV
but without significant bumps.

The possibility that the anomalies are due to annihilating or decaying DM
is very exciting, 
although standard astrophysical sources,
such as pulsars or nearby supernova remnants
\cite{Voelk95,2001A&A...368.1063Z,
Kobayashi:2003kp,Hooper:2008kg,Yuksel:2008rf,Profumo:2008ms,Ioka:2008cv,
Shaviv:2009bu,Malyshev:2009tw,Kawanaka:2009dk},
provide a viable explanation as well.

In order for DM annihilation to produce the necessary $e^+e^-$ flux,
the cross section must be 100 - 1000 times larger than the standard
value 
($\langle \sigma v \rangle \sim 3 \times 10^{-26}$ cm$^3$ s$^{-1}$)
inferred from the relic DM abundance. 
Boosts due to DM inhomogeneities
are expected to be $\lesssim 10$ \cite{Diemand:2008in} in the solar neighborhood,
and an additional mechanism, such as the Sommerfeld enhancement,
is necessary
\cite{Cirelli:2008id,ArkaniHamed:2008qn,Lattanzi:2008qa}.
The Sommerfeld enhancement can be further amplified in subhalos
due to their lower velocity dispersions \cite{Kuhlen2009}.
DM models, proposed to explain the ATIC and
PAMELA anomalies, include the lightest supersymmetric particles
\cite{Dimopoulos:1996gy,Bergstrom:2008gr,
Allahverdi:2008jm,Banks:2009rb,Cheung:2009qd,
Katz:2009qq},
Kaluza-Klein particles 
\cite{Servant:2002aq,Cheng:2002ej,Hooper:2009fj},
and various phenomenological scenarios 
\cite{Cholis:2008vb,Cholis:2008hb,Cirelli:2008pk,Cholis:2008qq,
Ponton:2008zv,Cholis:2008wq,Phalen:2009xw}
(for a recent review see, e.g., \cite{Hooper:2009zm,Bergstrom:2009ib})

In the presence of a large number of competing possibilities, one of the 
most important questions is
what properties of the $e^+e^-$ flux are model dependent and what properties are universal.
In this work we study the dependence of the $e^+e^-$ flux on
\bi
\item DM models,
\item the shape of DM host halo, and
\item the presence of subhalos.
\ei
The dependence on DM models and on the host halo density profile was recently investigated in \cite{Cholis:2008vb,Chen:2008fx,Meade:2009rb,Mardon:2009rc}.
One may notice that the flux is nearly independent of the density
profile and that the choice of DM model seems to affect the spectral
shape only at high energies near the break.

We show that for any DM model there exists an energy $E_* \lesssim \mdm$
such that the $e^+e^-$ flux has a universal behavior $F_{e^\pm}(E) \sim E^{-2}$ for energies \mbox{1 GeV $ \ll E \ll E_*$}.
In general, for $E \ll E_*$,
the index depends on the energy loss function, $\dot E = - b(E)$, for electrons propagating
in the interstellar medium (ISM)
\be
\lb{flux-index}
F_{e^\pm} \sim \frac{1}{b(E)}\qquad
E \ll E_*,
\ee
where $b(E) \sim  E^2$ for $E \gg$ 1 GeV. The behavior in Eq. (\ref{flux-index}) does not depend on the DM model,
but the value of $E_*$ is model dependent. 
Eq. (\ref{flux-index}) may be also modified at energies $E \sim 10$ GeV due to a dependence
on the DM density profile.

This paper is organized as follows: In Sec. \ref{sec:Green}, we
review the propagation of electrons and positrons in the ISM and
derive Eq. (\ref{flux-index}). In Sec. \ref{sec:Dark}, we study
the dependence of the spectral index on DM models and on the spatial
distribution of the DM. For our purposes, the most important
difference between DM models is the number of steps in the
annihilation-decay process, with more decay steps leading to a softer
$e^+e^-$ injection spectrum. However, propagation effects then ensure
that the final spectrum exhibits the universal behavior of Eq.
(\ref{flux-index}). Likewise, variations in the DM spatial
distribution also do not change this behavior. As we show, the spectral
slope is not affected by linear gradients in the source function, with
the leading order correction arising from second derivatives.  The Earth
is sufficiently far from the Galactic center such that the corresponding
effects are small, although the corrections may become important at $E
\sim 10$ GeV due to an increase in the characteristic diffusion distance.

A local clump of DM, on the other hand, may
result in a significant variation in the spectral index.
In Sec. \ref{sec:Clumps}, we investigate this possibility
using {\VLII} (hereafter VL2), one of the highest resolution numerical simulations of
the DM structure of a Milky-Way-scale halo \cite{Diemand:2008in,
  Kuhlen:2008aw, Madau:2008uz}. We find that on average there is one
$M>10^5 M_\odot$ clump within 3 kpc from Earth. Without Sommerfeld
enhancement, the likelihood of boosting the local flux by a
significant factor ($>$10) due to one such nearby subhalo is very low,
in agreement with previous findings
\cite{Diemand:2008in,Brun:2009aj}. Allowing for Sommerfeld
enhancement, however, the average luminosity from nearby clumps can
match (or even exceed) the smooth host halo contribution. We show
that, in general, the electron flux from a nearby DM clump is harder
than from the host halo (as also observed in
\cite{Pohl:2008gm,Hooper:2008kv,Brun:2009aj,Ilias}). In the DM
annihilation picture, the high energy ($\sim 600$ GeV) flux observed
by ATIC, PPB-BETS, and HESS is consistent with being produced by a
single DM clump, but in order to explain PAMELA the flux must be
dominated by the smooth host halo below energies of $\sim 100$ GeV
(another possibility to explain PAMELA is via a second, more distant,
large clump of DM as pointed out in \cite{Brun:2009aj}).  
Recent Fermi and HESS data are significantly smoother than the ATIC data
and are fitted better with the flux from the host halo without contributions 
from large subhalos.
We conclude
in Sec. \ref{sec:Concl} by summarizing which properties of the
$e^+e^-$ flux are model dependent and which are universal.

\section{Propagation of electrons}
\lb{sec:Green}

The propagation of electrons in the ISM is described by 
diffusion in the Galactic magnetic field and energy losses \cite{Ginzburg1964, Longair1992}.
At energies $E > 10$ GeV, the main losses are from synchrotron radiation and inverse
Compton scattering, which can be estimated as
\be
\dot E = - b_0 E^2,
\label{energy_losses}
\ee
where $b_0 = 1.6 \times 10^{-16}\; \text{GeV}^{-1} \text{s}^{-1}$ 
for the local densities of starlight, infrared, and CMB photons \cite{Porter:2005qx}
and a magnetic field of $3 \mu$G \cite{Longair1992}.
For relativistic electrons, the diffusion coefficient can be expressed as
\be
D(E) = D_0 \left(\frac{E}{1 \; {\rm GeV}}\right)^\dl.
\ee
In the examples, we will use $D_0 = 20$ pc$^2$ kyr$^{-1}$ and 
$\dl = 0.6$
\cite{Maurin:2001sj,Delahaye:2007fr,Strong:2007nh}.
We will also assume that the diffusion zone in the vertical direction (away from the
Galactic plane) is sufficiently large, $L \gtrsim 4$ kpc, such that 
the escape of electrons from the Galaxy can be neglected.
Models with a small diffusion zone, $L \lesssim 1$ kpc, tend to have exponential 
suppression of the flux from DM at energies $E \lesssim100$ GeV \cite{Chen:2008fx}, which would contradict the PAMELA data.
Our main qualitative conclusions will not depend on the choice
of parameters $b_0$, $D_0$, and $\dl$.
As a useful simplification, we will also assume that these parameters do not depend
on $\vx$.

The electron energy spectrum $f = \frac{dN}{dE\;dV}$ evolves according to the
diffusion-loss equation
\cite{Ginzburg1964, Longair1992}
\be
\lb{diff-loss}
\frac{\p f}{\p t} = 
\frac{\p }{\p E}  \text{\Large (}b (E) {f}  \text{\Large )}
+ D (E) \nabla^2 {f}
+ Q(\vx, E, t).
\ee
The general solution of this equation
is found in \cite{Ginzburg1964, Syrovatskii1959}.
If the source is constant in time, then $\p {f} / \p t = 0$ and the equation simplifies
to the usual diffusion equation.
If we introduce a new variable $\ld(E)$ such that
\be
\frac{d \ld}{d E} = - \frac{D(E)}{b(E)},
\ee
then
\be
\lb{diff-eq}
\frac{\p (b {f})}{\p \ld} 
- \nabla^2 (b {f})
= \frac{b(E)}{D(E)}Q(\vx, E).
\ee
The Green's function for the diffusion equation is
\be
G(\vx, \ld) = \frac{1}{(4\pi\ld)^{3/2}} e^{-\frac{\vx^2}{4\ld}},
\ee
and so the general solution to Eq. (\ref{diff-eq}) is given by
\bea
\nonumber
{f}(\vx, E) &=& \frac{1}{b(E)} \int d^3 x_0 \int_E^\infty dE_0\;
G\text{\large (}\vx - \vx_0, \ld(E,E_0) \text{\large )}\\
\lb{density-soln}
&& \cdot \; Q(\vx_0, E_0),
\eea
where
\be
\ld(E, E_0)   = \int_E^{E_0} \frac{D (E')dE'}{b (E')}.
\ee
Our approximation is valid if the characteristic propagation distance,
$x^2 = 6\ld(E, E_0)|_{E_0 \ra \infty}$,
is much smaller than the size of the vertical diffusion zone $L$.
For our choice of parameters, this gives
\be
E \gtrsim 
\left(
60\frac{\text{kpc}^2}{L^2}
\right)^\frac{1}{1-\dl} {\rm GeV}
\ee
If $L \sim 4$ kpc, then $E \gtrsim 30$ GeV.
For smaller energies there is an exponential suppression due to escape of 
electrons from the Galaxy.

For a homogeneous monochromatic source 
$Q(\vx_0, E_0) = Q_0 \dl(E - \mdm)$,
the solution to Eq. (\ref{density-soln}) is
\be
\lb{monochrom}
{f}(E) = \frac{Q_0}{b(E)},
\ee
which has the behavior (\ref{flux-index}) announced in the Introduction.
For energies $E \gg 1$ GeV, $b(E) \sim E^2$ (cf. Eq.~(\ref{energy_losses})) and therefore ${f}(E) \sim E^{-2}$.
One of the main purposes of the paper is to elucidate
how the index $n = -2$ changes for different DM models and
different DM distributions.

\section{Dark Matter in a Host Halo}
\lb{sec:Dark}

The source function of electrons and positrons coming from annihilating DM is
\be
\lb{DM_flux_annih}
Q (\vx, E) = \frac{1}{2}\frac{\rhodm^2}{\mdm^2} \bra \sigma v \ket \frac{dN_{\rm ann}}{dE},
\ee
where $\frac{\rhodm}{\mdm}$ is the DM number density, $\langle\sigma v \rangle$ is the thermally averaged annihilation cross-section,
and ${dN_{\rm ann}}/{dE}$ is the differential energy spectrum of electrons and positrons produced in a single annihilation event.
Here we assume that the DM particle is its own antiparticle, e.g.,
a Majorana fermion,
otherwise there is an additional factor of $1/2$.

As we discuss in more detail below, the cross section may exhibit a
velocity dependent boost factor 
$\bra \sm v \ket = \bra \sm v \ket_0 \; S(v)$, 
where $\bra \sm v \ket_0$ is the cross section at freeze-out
and $\displaystyle \lim_{v\to c} S = 1$. Generically, this enhancement
scales as $S \sim 1/v$, although resonances exist for certain
parameter choices, in which case $S \sim 1/v^2$
\cite{Lattanzi:2008qa}. Note that in either case the boost saturates
at small velocities.

DM halos in general are not isothermal and have lower
velocity dispersions, and hence higher boost factors, in their centers
\cite{Diemand:2004kx,Hoeft:2003ea,Dehnen:2005cu}. Given a density
profile, and assuming equilibrium and spherical symmetry, it is
straightforward to solve the Jeans' equation for the corresponding
velocity dispersion profile $\sigma_{\rm v}(r)$.  
For a Navarro, Frenk, and White (NFW) density
profile, the resulting velocity dispersions peak at about the scale
radius and decrease towards the center as $\sigma_{\rm v}(r) \sim
r^{1/2}$ \cite{Robertson:2009bh}. This is a limiting case
\cite{Tremaine:1993qb}, and all other profiles (e.g.  the Einasto
profile) have shallower velocity dispersion profiles.

The source function (\ref{DM_flux_annih}) can be split into a product of $\vx$-dependent and 
$E$-dependent functions
\be
\lb{source-split}
Q (\vx, E) = \kappa L(\vx) Q(E),
\ee
where 
\be
\kappa = \frac{\bra \sm v \ket_0}{2\mdm^2}.
\ee

The $\vx$-dependent part is the luminosity
\be
L(\vx) = \rhodm^2(\vx)\; S(v(\vx)),
\ee
and the $E$-dependent part is the injection spectrum
\be
Q(E) = \frac{d N_{\rm ann}}{d E} (E),
\ee
where we choose the normalization such that
$\int_0^\infty Q(E) dE$ is the total average number of electrons and positrons 
produced in an annihilation.

\subsection{Sommerfeld enhancement}

In the following we will assume the standard DM density
$\rhodm = 0.3$ GeV cm$^{-3}$ 
and a freeze-out cross section 
$\bra \sm v \ket_0 = 3 \times 10^{-26}$ cm$^3$ s$^{-1}$ .
This cross section is too small to give a sufficient annihilation rate for the
PAMELA and ATIC anomalies, and hence a boost factor of 100 - 1000 is necessary.
Recent Fermi data are consistent with this conclusion.
If we assume that the PAMELA anomaly is due to dark matter,
then the absence of steplike features in Fermi data below the cutoff
requires $M_{\rm DM} \gtrsim 1$ TeV.
Consequently the same boost factor 100 - 1000 is required to explain PAMELA.

An elegant way to increase the annihilation cross section without
affecting the relic abundance, is to assume the existence of a new
light force carrier $\phi$ in the dark sector, $m_\phi \sim
\mathcal{O}$(GeV). This additional force results in a Sommerfeld
enhancement \cite{ArkaniHamed:2008qn}, which is unimportant at
freeze-out when particles are close to relativistic, but can
significantly boost annihilation rates today.  There may be further
enhancement, if the annihilation proceeds through the formation of
\mbox{WIMPonium} \cite{Pospelov:2008jd}, a meta-stable bound state of
two DM particles. This process is analogous to the annihilation of
electrons and positrons through positronium, which is dominant at low
velocities. For some parameters in the DM sector, it is possible to
radiatively create \mbox{WIMPonium} by emitting a $\phi$ particle, and
in this case the \mbox{WIMPonium} annihilation channel dominates over
immediate annihilation.

We now demonstrate that in every model with 
parameters that do not allow for \mbox{WIMPonium} creation,
there exists an upper bound,
\be
\lb{boost_limit}
S \lesssim \frac{2\pi}{\sqrt{v}},
\ee
on the nonresonant boost factor,
where $v$ is the relative velocity of the particles.

Generic (nonresonant) Sommerfeld enhancement is \cite{ArkaniHamed:2008qn}
\be
S(v) = \frac{\pi \al}{v}.
\ee
This enhancement factor saturates when the deBroglie wavelength
of the DM particles becomes equal to the force range
$\frac{1}{\mdm v} \sim \frac{1}{m_\phi}$, thus
\be
S \lesssim \frac{\pi \al \mdm}{m_\phi}.
\ee
\mbox{WIMPonium} cannot be created if the corresponding binding energy is smaller
than the mass of the mediator
\be
\frac{1}{4}\al^2 \mdm \lesssim m_\phi.
\ee
Collecting these pieces together, we find
\be
\frac{\pi \al}{v} \lesssim  \frac{\pi \al \mdm}{m_\phi} \lesssim  \frac{4\pi}{\al}, 
\ee
and thus
\be
\al \lesssim 2\sqrt{v},
\ee
which gives the bound (\ref{boost_limit}). 
In any particular model the precise value of the numeric coefficient in (\ref{boost_limit})
may be different, but the parametric form should be the same 
(unless there are some additional reasons that prevent WIMPonium creation).

At the location of the Sun (8 kpc), the velocity dispersion of the DM particles is $\sigma_{\rm v} \approx 200$ km/s \cite{Kuhlen2009}. Taking this as a proxy for the relative velocities of annihilating DM particles, we find
\be
S(200\;{\rm km/s}) \lesssim 250.
\ee
Thus, the local Sommerfeld enhancement, $S$(200 km/s), cannot exceed $\sim\,$250 without taking into account
\mbox{WIMPonium} creation or resonance effects.
If we allow for \mbox{WIMPonium} creation ($m_\phi < {1}/{4}\al^2 \mdm$),
then the annihilation cross section increases by an additional factor of 7 (3) for fermionic (bosonic) DM particles \cite{Pospelov:2008jd}, i.e.,
for fermionic DM particles the enhancement factor is
\be
S(v) \approx 20 \frac{\al}{v}.
\ee

In order to parameterize the Sommerfeld enhancement as a function of velocity,
we need to specify the normalization $S_0$ and the saturation velocity $v_{\rm min}$.
Then, for small velocities ($v \ll \al$),
\be
S(v) = S_0 \frac{v_0}{v + v_{\rm min}}.
\ee

We take $v_0$ to be the local velocity dispersion, $v_0=200$ km/s.
In the examples we discuss below, we set $v_{\rm min} = 3$ km/s and
$S_0 = 600$.
These parameters are rather generic and can be obtained, for instance,
in a fermionic DM model with a vector boson mediator with
$\al = 1/50$ and $m_\phi / \mdm = 10^{-5}$.
The small mass ratio is necessary to have a small saturation velocity,
$\frac{v_{\rm min}}{c} \sim \frac{m_\phi}{\mdm}$.
For small $v_{\rm min}$ the role of small subhalos is greater since 
their velocity dispersion is smaller.
For larger $v_{\rm min}$ the luminosity of small subhalos becomes smaller
and the probability to have an observable effect is reduced.

For $\mdm \sim 1$ TeV we have $m_\phi \sim 10$ MeV $> 2 m_{e^\pm}$,
i.e. this model is kinematically viable.
In building a realistic model, one has to check various
constraints from diffuse gamma rays and neutrino fluxes
\cite{Beacom:2006tt,Liu:2008ci,Meade:2009rb,Mardon:2009rc,Barger:2009yt}.
However,
instead of trying to find a specific model that satisfies all current
observational data, we now discuss some general properties of
the $e^+e^-$ flux from annihilating DM.

\subsection{DM model dependence}
\lb{sec:DM-models}

For a homogeneous monochromatic source,
the electron spectrum was derived in Eq. (\ref{monochrom}).
In general, the spectrum can be approximated as
${f} \sim E^{\,n}$ with an energy dependent index $n(E)$.
Since at low energies we expect ${f}(E) \sim 1 / b(E)$,
the variation of the index will be defined as
\be
\dl n \equiv \frac{d \log b(E) {f}(E)}{d \log E}.
\ee
For a homogeneous source,
the electron density is
\be
{f} (E) 
= \frac{1}{b(E)}\int_E^{\mdm} Q(E_0) dE_0.
\ee
If the injection spectrum has the form of a power law
\be
Q(E) \sim E^{\al}\qquad \al > -1,
\ee
then the integral is saturated at $E _0 \sim \mdm$ and the variation of $n$ is small
for $E \ll \mdm$.

We define the break energy $E_*$ by the condition
\be
\lb{break-cond}
\dl n(E_*) = -1.
\ee
The flux $F_{e^\pm} \sim E^{-3}$ near $E_*$,
which is approximately the same scaling as the backgrounds 
\cite{:2008zzr,Moskalenko:1997gh}.
Thus the ratio $\frac{F_{e^\pm}}{F_{\rm backgr}}$ is maximal near $E_*$.

As an example, we consider DM annihilation followed by a chain
of $(k - 1)$ two-body decays
\be
2\chi \;\ra\; 2 \phi_1 
\;\ra\; 4 \phi_2 
\;\ra\; \ldots 
\;\ra\; 2^{k - 1} e^+ + 2^{k - 1} e^-
\ee
We assume that $m_{i+1} \ll m_i$, i.e. the decay products of $\phi_i$ are 
relativistic in the rest frame of $\phi_i$.
The spectrum of $\phi_i$ particles is \cite{Mardon:2009rc}
\be
\lb{DMmodels}
\ba{lll}
Q_1(E) &=& 2 \dl(M_\chi - E) \\
Q_i (E) &=& \frac{2^i}{(i - 2)!}\left(\log \frac{\mdm}{E} \right)^{i - 2}
\qquad i > 1,
\ea
\ee
where we include the multiplicity $2^i$ in the definition of the source functions
$Q_i(E)$.

\begin{figure}[t] 
\begin{center}
\epsfig{figure = 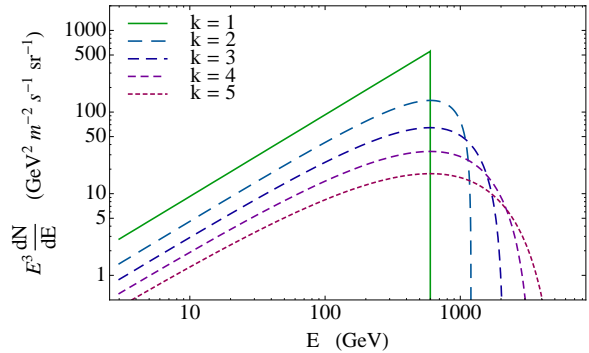,scale=0.4}
\end{center}
\vspace{-2mm}
\noindent
\caption{\small The $e^+e^-$ flux depending on the number of steps in the toy model
DM annihilation (see text).
In order to keep the maximum values of $E^3 F$ at $E_* = 600$ GeV,
we choose the DM mass $\mdm \sim e^{\frac{k - 1}{2}} E_*$.
Apart from overall normalization, the spectra look similar to each other for $E \ll E_*$.
Model dependence shows up for energies $E \gtrsim E_*$. 
}
\label{ModelDep}
\vspace{2mm}
\end{figure}

The DM particle mass is not directly observed, rather the experiments measure the peak of the bump, i.e. $E_*$, which is a function of $\mdm$.
Motivated by the ATIC data we fix $E_* = 600$ GeV
and find the necessary $\mdm$ in order to satisfy condition (\ref{break-cond})
\be
\ba{c c c c c c c c}
\frac{\mdm}{E_*} & = & 1, & 2, & 3.5, & 6,  & 10 & \ldots \\
k &=& 1, & 2, & 3, & 4, & 5 & \ldots \; .
\ea
\ee
For large $k$, the ratio grows exponentially,
$
\frac{\mdm}{E_*} \approx e^\frac{k - 1}{2}.
$
The corresponding fluxes for $k \leq 5$ are shown in
Fig.~\ref{ModelDep}.  At energies $E \ll 600$ GeV the spectra
look very similar to each other, indicating model independence. At
higher energies, on the other hand, the spectral shape is sensitive
to the value of $k$, i.e. it depends on the DM model.

\subsection{Profile dependence}

In this section we calculate 
the variation of the index due to a coordinate dependence of the luminosity.
In order to find the spectrum of electrons
coming from DM annihilating in the host halo,
we take the source function
(\ref{source-split}) 
and substitute it in Eq. (\ref{density-soln}).
At the observer's position $\vx = 0$, we get
\bea
\nonumber
b(E){f}(E) &=& \int_E^\infty dE_0 \;\kappa\; Q(E_0)\\
&& \cdot\int d^3 x_0 \; \frac{1}{(4\pi\ld)^{3/2}} e^{-\frac{\vx_0^2}{4\ld}} \; L(\vx_0)
\eea
In the previous section we considered the variation of the index due to the lower limit 
of the $E_0$ integration. Now the integral over $x_0$ is a function of $\ld(E, E_0)$ and this introduces an additional $E$ dependence.

The resulting variation is
\be
\lb{coord-varn}
\dl n \sim
\frac{d \log \ld}{d \log E}\;\; \frac{\bra \frac{\vx^2}{2 \ld} L \ket - 3 \bra L \ket}{\bra L \ket},
\ee
where 
$\frac{d \log \ld}{d \log E} \sim (\dl -1)$ and 
the average of a function $f(x)$ is defined as
\be
\bra f \ket = \int d^3 x \frac{1}{(4\pi\ld)^{3/2}} e^{-\frac{\vx_0^2}{4\ld}} \; f(x).
\ee
In particular, $\bra \vx^2 \ket = 6 \ld$. Thus, for a uniform source, $\dl n = 0$.
For a linear gradient $\dl L = b_i x^i$, 
$\bra \dl L \ket = \bra \frac{\vx^2}{2\ld} \dl L \ket = 0$.
Consequently, the first correction to $n$ follows from the second derivative 
of $L(\vx)$.

\begin{figure}[t] 
\begin{center}
\epsfig{figure = 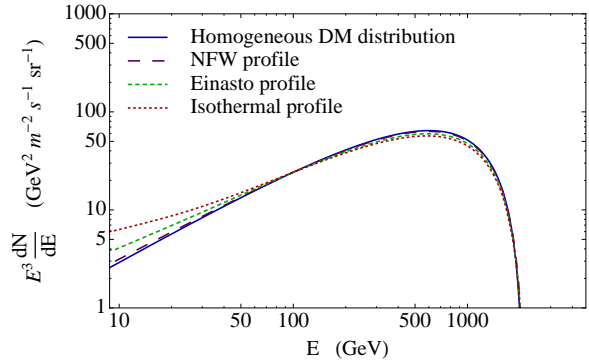,scale=0.4}
\end{center}
\vspace{-2mm}
\noindent
\caption{\small The dependence of $e^+e^-$ flux on the DM halo profile.
We use NFW \cite{Navarro:1996gj}, 
Einasto ($\al = 0.17$) \cite{Navarro:2003ew}, 
and Isothermal \cite{Bahcall:1980fb,Kamionkowski:1997xg}
profiles.
The difference from the homogeneous distribution is proportional
to ${x^2_{\rm diff}(E)} / { r_0^2 }$, where $r_0 = 8$ kpc is the distance from the center of the Galaxy.
For small energies the characteristic diffusion distance increases and 
the corrections are more significant.
Similar results were obtained in \cite{Hooper:2004bq}
for $\mdm = 300$ GeV.
}
\label{DMhalo}
\vspace{2mm}
\end{figure}

For slowly varying $L(\vx)$,
\be
\lb{var-n}
\dl n \sim - (1 - \dl) \;\ld \frac{\nabla^2 L}{L}.
\ee
At the position of the Earth, the luminosity can be approximated as
\be
L(r) = \rhodm^2 S \sim r^{-\al}.
\ee
Consequently
\be
\nabla^2 L(r) = \frac{1}{r^2}\p_r r^2 \p_r\; (\rhodm^2 S) \sim \frac{\al(\al - 1)}{r^2} L(r).
\ee
Substituting this expression in Eq. (\ref{var-n}), we get
\be
\dl n \sim - (1 - \dl)\al(\al - 1) \frac{\ld}{r^2_0},
\ee
where $r_0 \approx 8$ kpc is the distance from the center of the Galaxy to Earth.
Since $\ld \ll r_0^2$, this variation is generally small (Fig. \ref{DMhalo}).

The situation is different when the density inhomogeneity is due to a discrete subhalo. For a DM clump 
with size $l_{\rm cl} \ll \sqrt{\ld}$ we can neglect the first term
in the numerator of Eq.~(\ref{coord-varn}).
The correction to the index then becomes positive and $\mathcal{O}$(1),
$\dl n \sim - \frac{d \log \ld}{d \log E} \approx (1 - \dl)$.
This leads to a harder electron spectrum, 
and the index is no longer independent of the DM model and the properties of ISM.
We consider some examples in the next section.

\section{Clumps of Dark Matter}
\lb{sec:Clumps}

In this section, we describe the properties of subhalos using the
results of the {\VLII}
(VL2) simulation and study their
influence on the electron flux spectrum.

\subsection{Via Lactea subhalos}

\begin{figure}[th] 
\begin{center}
\epsfig{figure = 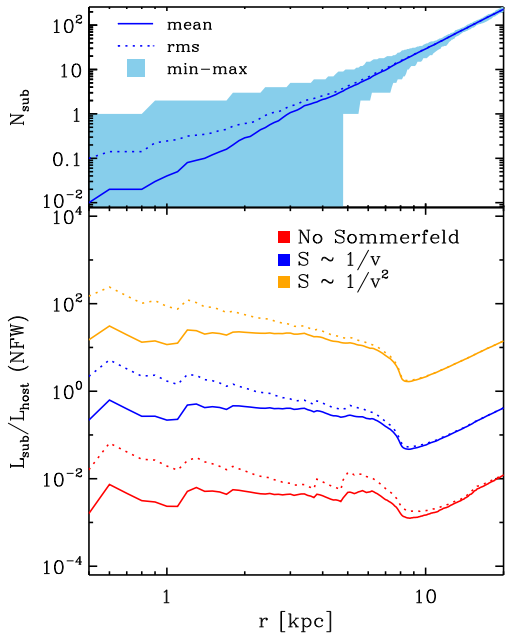,scale=0.7}
\end{center}
\vspace{-2mm}
\noindent
\caption{\small Some properties of DM subhalos using results of {\VLII} simulation
\cite{Diemand:2008in}.
The upper plot shows the expected number of subhalos within a distance $r$ from the Earth.
In the lower plot we show the ratio of luminosities of all subhalos within $r$ to the 
total luminosity of the host halo in the solid sphere of radius $r$ around the Earth.
We plot the ratio of luminosities without Sommerfeld enhancement, with generic $S \sim 1 / v$ enhancement, and with resonant $S \sim 1 / v^2$ enhancement. These curves only account for subhalos resolved in the {\VLII} simulation, $M_{\rm sub} \gtrsim 10^5 M_\odot$, and should be viewed as lower limits.
}
\label{ViaLactea}
\vspace{2mm}
\end{figure}

With a particle mass of 4000 $M_\odot$, VL2 can resolve subhalos down
to $\sim 10^5 M_\odot$.  For our purposes, the important
subhalo properties are their size, their number density, and their
luminosity. Since the annihilation rate scales as $\rho^2 S$, it is
strongly peaked toward the centers of subhalos.  For an NFW profile,
$\sim 90\%$ of the total annihilation luminosity originates from
within the scale radius, which is much smaller than the typical
propagation distance of $\sim$ 1 kpc.  Consequently, we will treat the
subhalos as point sources.

We determined the properties of the local subhalo population from the
simulation by randomly placing 100 sample spheres of radius 20 kpc,
each at a distance of 8 kpc from the host halo center.  For each
sample sphere, we identified all subhalos whose centers fall within
the sphere. The top panel of Fig.~\ref{ViaLactea} shows the mean,
root mean square, minimum, and maximum number of subhalos over the
ensemble, as a function of radius within the sample sphere. To a good
approximation the mean number of subhalos grows as $r^3$, and hence we
will assume that the subhalo number density is constant. On average
there is one subhalo within $r \approx 3$ kpc. Note that these results
refer solely to the portion of the substructure hierarchy that is
resolved in the VL2 simulation, i.e. $M_{\rm sub}>10^5 M_\odot$. The
subhalo mass function has been measured in simulations to be a simple
power law $dn/dM \sim M^{-\al}$ over 4 - 5 decades of subhalo mass
\cite{Diemand:2008in,Springel:2008cc}, with a logarithmic slope $\al
\approx 1.9$ independent of distance from the host halo center. 
This implies that the local density of subhalos with mass larger than $M$
may be extrapolated to $M < 10^5 M_\odot$ as
\be
n_{\rm sub}(M) \approx 9 \times 10^{-3} \; {\rm kpc}^{-3} \left( \frac{M}{10^5 M_\odot} \right)^{-0.9}.
\label{subhalo_abundance}
\ee

In the bottom panel of Fig.~\ref{ViaLactea} we plot the mean and the
root mean square of $L_{\rm sub}/L_{\rm host}$, the ratio of the
luminosity from all VL2-resolved subhalos to the smooth host halo
luminosity, as a function of the enclosed radius within the sample
spheres. We assigned a luminosity to each subhalo by assuming NFW
density and velocity dispersion profiles that are matched to the
values of $V_{\rm max}$ and $r_{\rm Vmax}$ as measured in the VL2
simulation. $V_{\rm max}$ is the peak of the circular velocity curve
$v_c^2(r)=GM(<r)/r$, and $r_{\rm Vmax}$ is the radius at which this
peak occurs, and both quantities are robustly determined in the
numerical simulation. We have checked that our results do not change
qualitatively if an Einasto profile is assumed instead.

We compare three different models: one with $S=1$ (no Sommerfeld
enhancement), one with $S \sim 1/v$, and one with $S \sim 1/v^2$.
Without Sommerfeld enhancement, the average contribution to the total
luminosity from $M>10^5 M_\odot$ subhalos is negligible, in agreement
with previous findings \cite{Brun:2009aj}. In the $S \sim 1/v$ model,
however, this subhalo contribution is already as large as that from
the smooth host halo, and in the $S \sim 1/v^2$ case it dominates by
close to 2 orders of magnitude. We caution that these are ensemble
averaged ratios: even at a distance of 3 kpc only 70\% of the sample
spheres contain one or more $M>10^5 M_\odot$ subhalo, and at 2 (1) kpc
this fraction drops to 24\% (3\%). The situation for models with
Sommerfeld enhancement can be summarized as follows:
\begin{itemize}
  \setlength{\itemsep}{1pt}
  \setlength{\parsep}{0pt}
\item[i)] when one or more $M>10^5 M_\odot$ subhalos happen to lie
  within the electron diffusion region, their combined flux dominates
  that of the smooth host halo;
\item[ii)]the probability of finding one or more such subhalos remains below 50\% out to 2.6 kpc, and a mean occupancy of one such subhalo is reached only at 3 kpc;
\item[iii)] the ensemble averaged expectation value of $L_{\rm
    sub}/L_{\rm host}$ is unity for $1/v$ models ($M>10^5  M_\odot$ subhalos are about as
  important as the host halo) and $\sim 100$ for $1/v^2$ models
  ($M>10^5  M_\odot$ subhalos are much more important than the host halo).
\end{itemize}
These results suggest that if a model relies on the Sommerfeld effect
to explain the high energy cosmic ray anomalies, then there is a
non-negligible probability of an individual DM subhalo affecting the
high energy electron flux.

We stress once more that these results refer merely to the subhalos
resolved in VL2. In the cold dark matter picture of structure
formation, the clumping of DM should continue far beyond the
artificial resolution limit of current state-of-the-art numerical
simulations like VL2. The total luminosity from all subhalos below the
simulation's resolution limit may very well dominate the local pair
production rate from DM annihilation, if the dense cores of such low
mass subhalos are able to withstand the disruptive forces from tidal
interactions with the DM host halo and with stars and molecular clouds
in the Galactic disk \cite{Goerdt:2006hp}. However, the diffusive
nature of the electron propagation would average out the contribution
from subhalos with an $\mathcal{O}(1)$ probability of lying within the
diffusion distance of $\sim$1 kpc, rendering their contribution
indistinguishable from the homogeneous host halo source. 
Let us define the characteristic mass scale $M_1$ such that
\be
\lb{norm_dens}
\frac{4\pi}{3}x_{\rm diff}^3 \int_{M_1}^\infty dn_{\rm sub} = 1.
\ee
where $x_{\rm diff} \sim 1$ kpc in our case.
For the local subhalo abundance of Eq.~(\ref{subhalo_abundance}), 
$M_{\rm 1} = 2,600\,M_\odot$.
Any additional
features in the electron spectrum from annihilating DM must arise from
a small number of large subhalos, $M > M_1$, within the diffusion region.

\subsection{The flux from a local clump}
\lb{sec:clump_flux}

\begin{figure}[t] 
\begin{center}
\epsfig{figure = 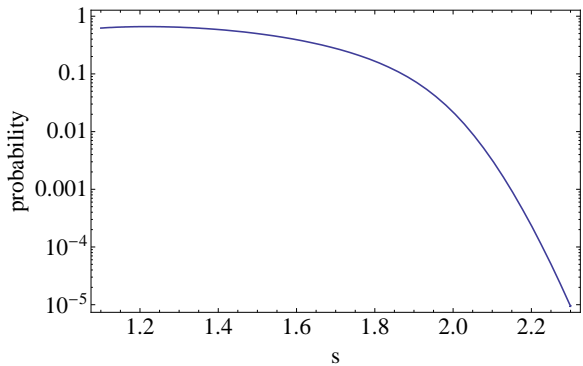,scale=0.4}
\end{center}
\vspace{-2mm}
\noindent
\caption{\small 
The probability to observe an order one feature at high energies from
a clump of dark matter
versus the index $s$ of the luminosity function [Eq. (\ref{lum_index})].
The break at $s = 2$ corresponds to equipartition of DM luminosity.
For $s < 2$ the luminosity is saturated by large clumps, 
for $s > 2$ the luminosity is saturated by small clumps.
The overall normalization is model dependent.
}
\label{ClumpProb}
\vspace{2mm}
\end{figure}

In this section we study an example of the flux from a single DM subhalo 
in an $S \sim 1/v$ model
and compare it with the flux from the host halo.

At first, let us estimate the probability 
to see a feature from a single DM clump
relative to the flux from the host halo and the averaged flux from minihalos (defined to have \mbox{$M<M_1$}).
Effectively, the answer to this question depends on only one parameter,
the logarithmic scaling $s$ of the subhalo number density with respect to the luminosity
\be
\lb{lum_index}
\frac{d n_{\rm sub}}{dL} \sim L^{-s}.
\ee
Qualitatively, if $s > 2$ then the integrated luminosity is saturated
by small clumps and the probability to have a significant flux from a
large clump is negligible.
If $s < 2$, then the luminosity is saturated by the large clumps and 
the probability to see an additional feature is significant.

Let us estimate the probability of an $\mathcal{O}(1)$ feature for the
equipartition distribution, $s = 2$.  In this case $n_{\rm sub}(L)
\sim L^{-1}$, i.e.  subhalos with 10 times larger luminosity have 10
times smaller density.  
Denote by $L_1$ the luminosity corresponding
to a subhalo of mass $M_1$.
Suppose that there are $p$ decades of subhalos below $M_1 \sim 10^3
M_\odot$, which in the equipartition case each contribute an equal
amount to the total luminosity.  For a minimum DM clump mass of
$10^{-12} - 10^{-4} M_\odot$ \cite{Profumo:2006bv,Bringmann:2009vf} we
can roughly estimate $p = 7 - 15$.

\begin{figure}[t] 
\begin{center}
\epsfig{figure = 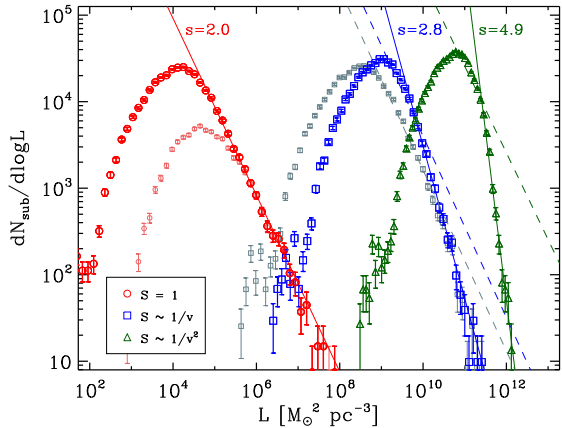,scale=0.55}
\end{center}
\vspace{-2mm}
\noindent
\caption{\small The luminosity function from VL2 subhalos, assuming
  NFW density and velocity dispersion profiles for the same three
  models as in Fig.~\ref{ViaLactea}. Without Sommerfeld enhancement
  (circles) the luminosity function is equipartioned ($s=2$) down to
  the simulation's completeness limit, \mbox{$L \sim 10^4 M_\odot^2$
    pc$^{-3}$}. For comparison we also show the distribution when the
  sample is restricted to subhalos with more than 250 particles ($M >
  10^6 M_\odot$) [small circles]. In the $S \sim 1/v$ Sommerfeld case
  (squares), the luminosity function is steeper ($s=2.8$) above the
  saturation luminosity of $L_{\rm sat} \sim 2 \times 10^9 M_\odot^2$
  pc$^{-3}$. Below saturation the distribution is expected to flatten
  to $s=2$ (dashed lines), and indeed this behavior is clearly seen
  for a model with a higher saturation luminosity of $L_{\rm sat} \sim
  3 \times 10^{10} M_\odot^2$ pc$^{-3}$ corresponding to $m_\phi/\mdm
  = 5 \times 10^{-5}$ (small squares). For $S \sim 1/v^2$ (triangles)
  the luminosity function above $L_{\rm sat}=10^{11} M_\odot^2$
  pc$^{-3}$ is even steeper ($s=4.9$).}
\label{LF}
\vspace{2mm}
\end{figure}

The total luminosity of subhalos with $L < L_1$ is then
$L_{\rm minihalos} = p L_1$.
For the purpose of estimation, suppose that the luminosity of 
the host halo is smaller than $L_{\rm minihalos}$;
then we should expect to observe an $\mathcal{O}(1)$ feature if 
there is a subhalo with luminosity $L_p = p L_1$ within the diffusion distance.
The probability of such a subhalo is 
\be
\pi_p \sim \frac{n_{\rm sub}(L_p)}{n_{\rm sub}(L_1)} = \frac{1}{p} \sim 0.1,
\ee
and is independent of the value of $L_1$ and of the details of the
propagation model.

For general $s$, the luminosity of a subhalo that is able to produce a significant
feature is
\bea
\nonumber
L_{p}(s) &\sim& \int_{L_{\rm min}}^{L_1} dL\: L\: \frac{dn_{\rm sub}}{d L} \\
&\sim& \frac{s - 1}{s - 2} 
\left[
\left(\frac{L_{1}}{L_{\rm min}}\right)^{s - 2} - 1
\right] L_1.
\eea
The luminosity $L_1$ is defined as
$\frac{4\pi}{ 3} x_{\rm diff}^3 \int_{L_1}^\infty d n_{\rm sub} = 1$
and $x_{\rm diff} \sim 1$ kpc in our case.
The corresponding density is
\be
n_{\rm sub}(L_p) \sim L_p^{1-s}
\ee
and the probability becomes
\be
\pi(s) \sim \frac{n_{\rm sub}(L_p)}{n_{\rm sub}(L_1)} =
\left( \frac{L_p(s)}{L_1} \right)^{1 - s}
\ee
The normalization of this probability is model dependent.
In particular, it depends on the ratio of the host
halo luminosity to that in the subhalos.
The shape of $\pi(s)$, however, is universal.
It has a break at $s = 2$ with a flat probability for luminosities
saturated by large clumps, $s < 2$, and a rapidly decaying 
probability for luminosity saturated by small clumps,
$s > 2$.
The function $\pi(s)$ for $L_1 / L_{\rm min} \sim 10^{10}$
is presented in Fig. \ref{ClumpProb}.

\begin{figure}[t] 
\begin{center}
\epsfig{figure = 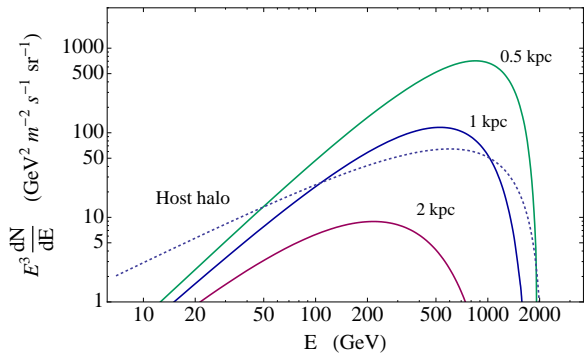,scale=0.4}
\end{center}
\vspace{-2mm}
\noindent
\caption{\small The flux of electrons from a typical DM 
subhalo at distances $r = 0.5,\: 1,\: 2$ kpc from Earth versus the flux from
the DM annihilation in the host halo.
The flux from a subhalo is significant for $r \lesssim 1$ kpc and
the corresponding spectrum is
harder than the spectrum from the host halo.
We use $S \sim 1 / v$ Sommerfeld enhancement of annihilation,
$\mdm = 2.1$ TeV, $m_\phi / \mdm = 10^{-5}$, $\alpha = 1/50$,
and $S_0 = 600$.
The subhalo luminosity is given in Eq. (\ref{subh-lum}).
}
\label{ClumpsVsHalo}
\vspace{2mm}
\end{figure}

In Fig.~\ref{LF} we show an empirical determination of the subhalo
luminosity function $dN_{\rm sub}/dL$ from the VL2 subhalos for the
same three models we considered in Fig.~\ref{ViaLactea}. As before,
the subhalo luminosities are estimated using an NFW density and
velocity dispersion profile, and we use the $S \sim 1/v$ model
normalized to $S_0=S(200\; \text{km/s})=600$. We include all subhalos
within the host's virial radius $r_{200}=402$ kpc. In the case without
Sommerfeld enhancment we find $s=2$ (equipartition) down to the
luminosity completeness limit of \mbox{$L \sim 10^4
  M_\odot^2$ pc$^{-3}$}. The turnover at smaller luminosities is due
to the fact that we do not resolve halos with $M < 10^5 M_\odot$ and
correspondingly low luminosities. When only subhalos with more than
250 particles ($M > 10^6 M_\odot$) are included, then the
distribution departs from a power law at roughly 10 times higher
luminosities.

\begin{figure}[th] 
\begin{center}
\epsfig{figure = 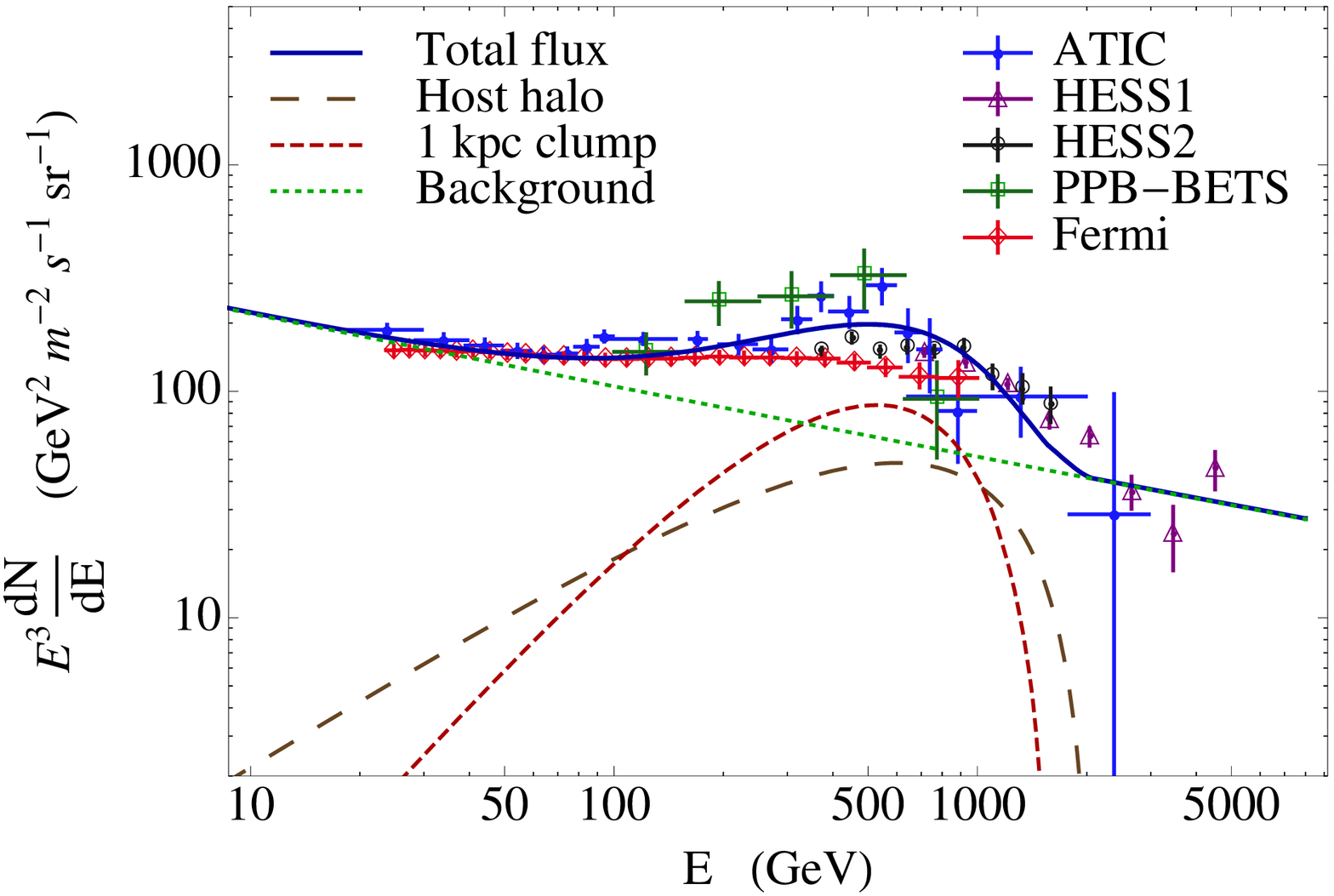,scale=0.4}
\hspace{3mm}
\epsfig{figure = 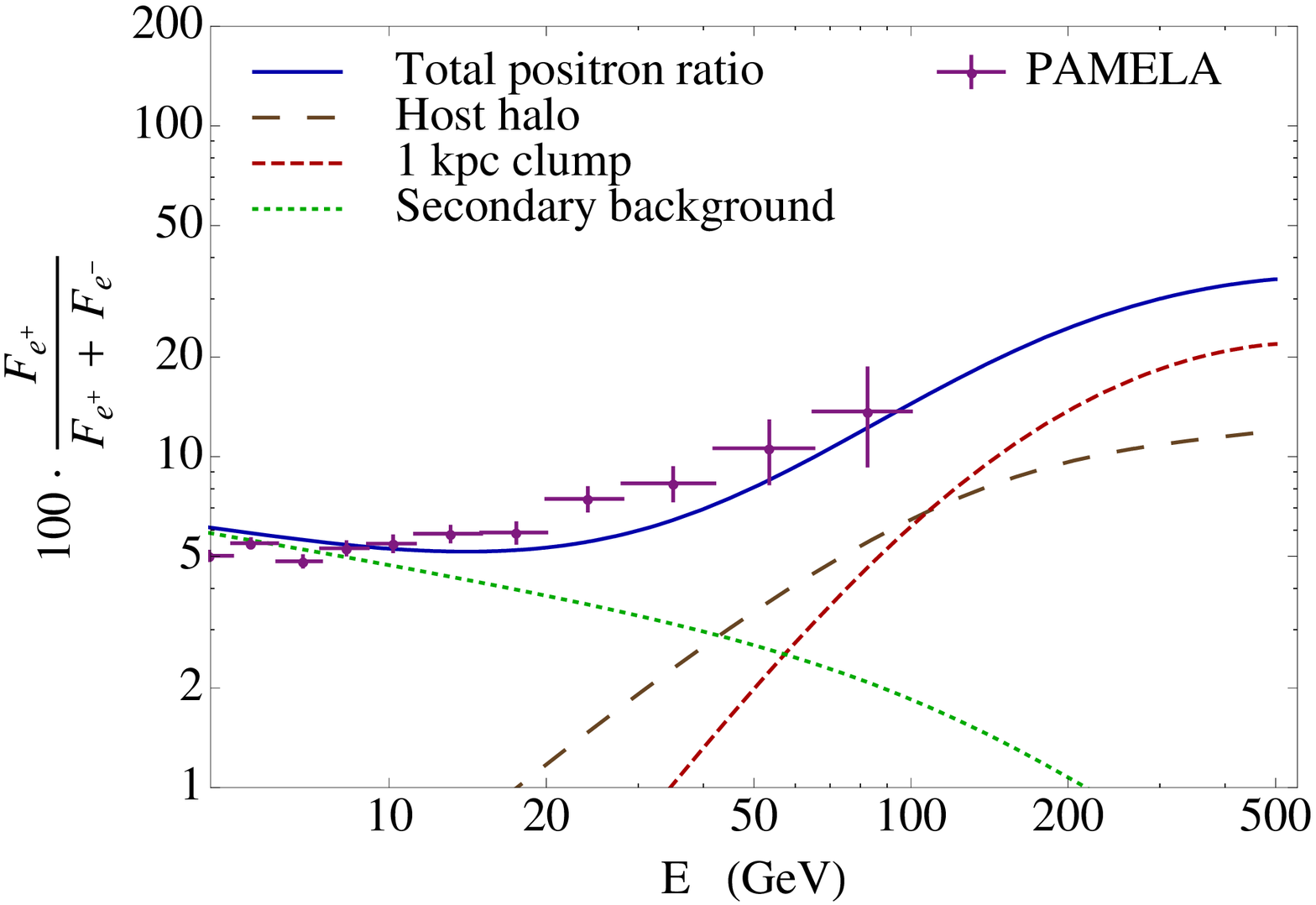,scale=0.4}
\end{center}
\vspace{-2mm}
\noindent
\caption{\small 
Electron and positron flux from annihilating DM 
in the host halo plus a 1 kpc clump
shown in Fig. \ref{ClumpsVsHalo}.
The hard spectrum of the flux from the subhalo is consistent with the ATIC bump
around 600 GeV, but, in order to fit the PAMELA data, we need the softer flux from
the host halo at energies $E < 100$ GeV.
We use the primary electron background $\sim E^{-3.3}$ and 
secondary electron and positron backgrounds
$\sim E^{-3.6}$.
The parameters of the 
DM model are the same as in Fig. \ref{ClumpsVsHalo}.
}
\label{1kpcAticPam}
\vspace{2mm}
\end{figure}

\begin{figure}[th] 
\begin{center}
\epsfig{figure = 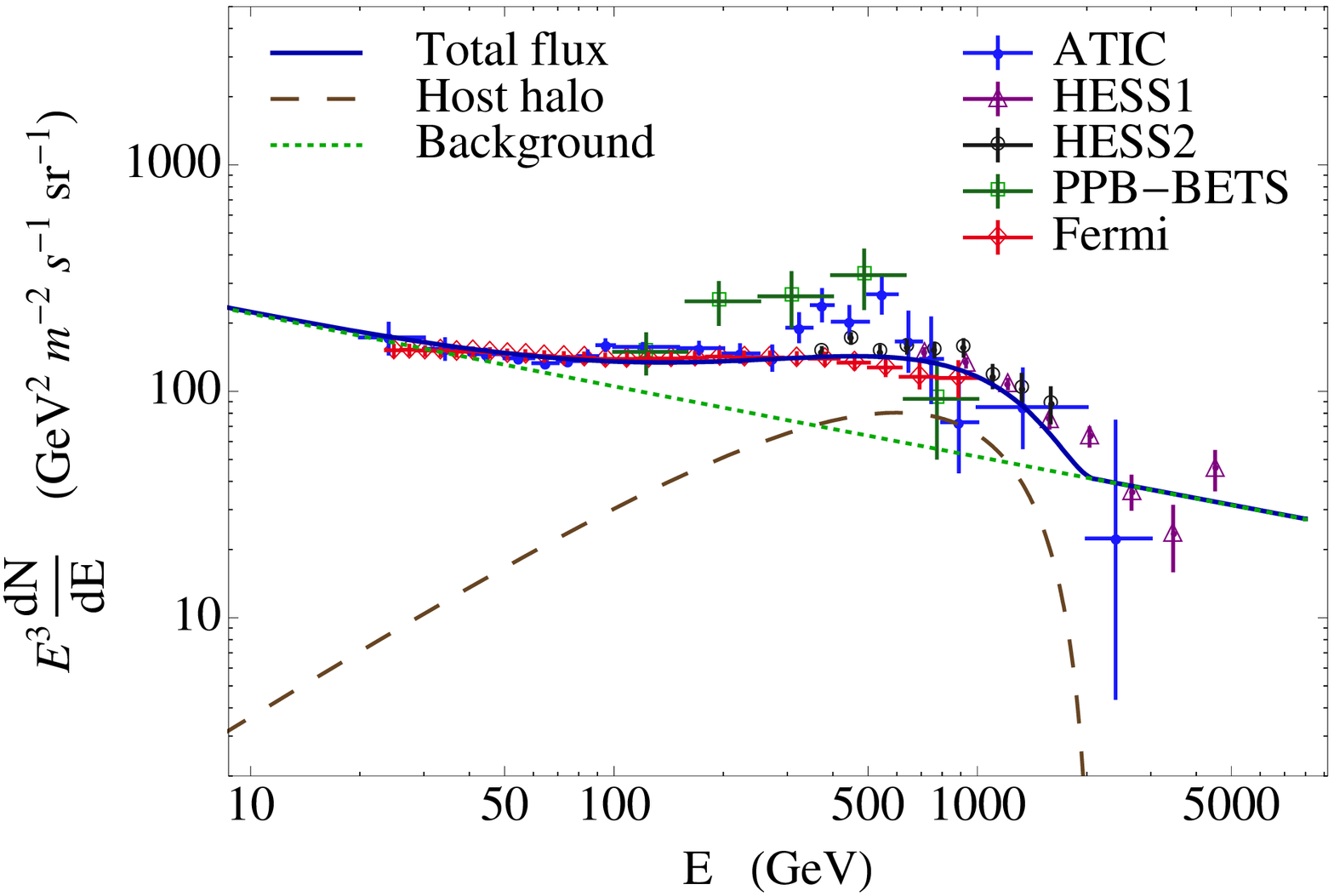,scale=0.4}
\hspace{3mm}
\epsfig{figure = 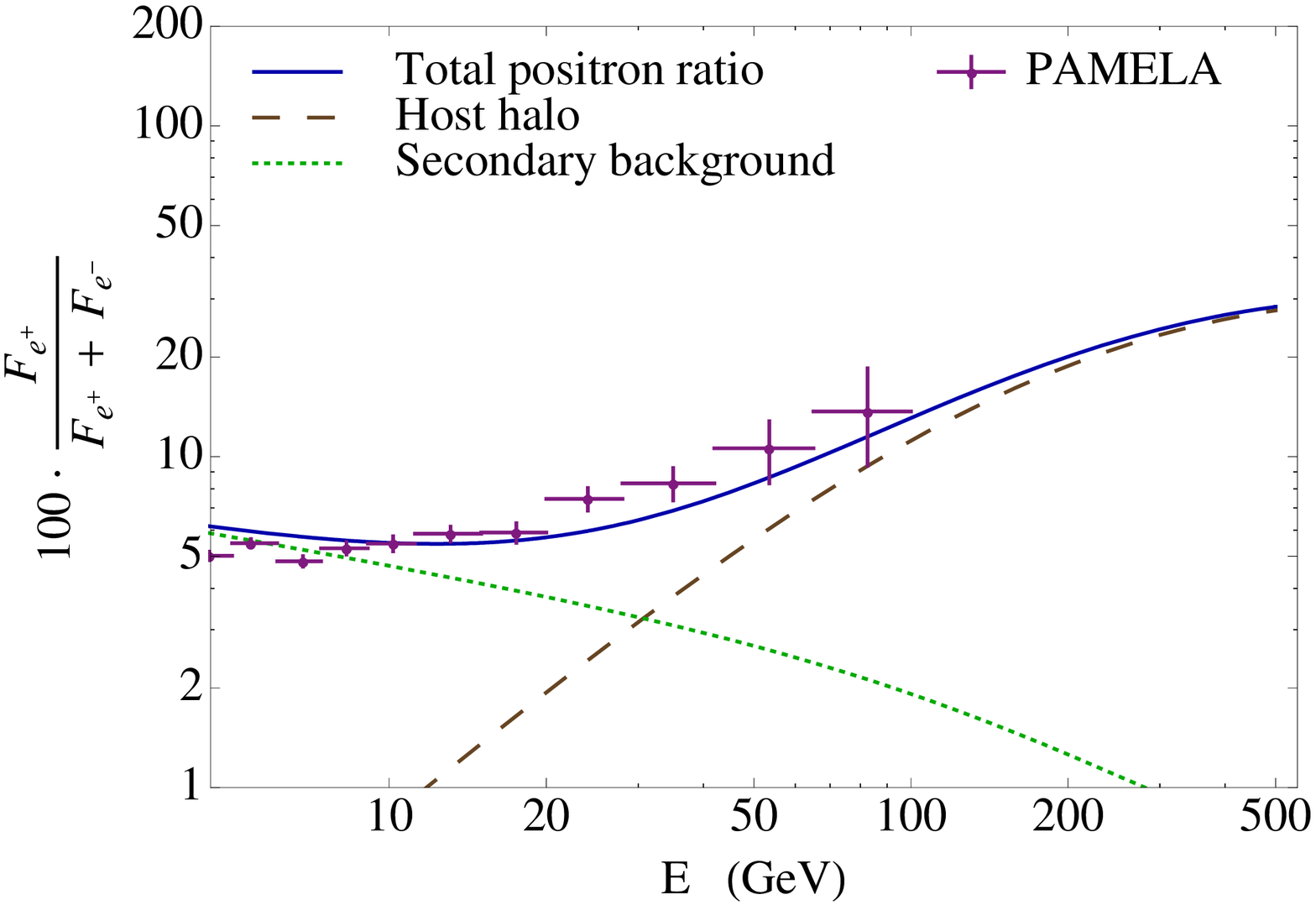,scale=0.4}
\end{center}
\vspace{-2mm}
\noindent
\caption{\small 
The same as in Fig. \ref{1kpcAticPam} but without the contribution from the 
subhalo and with a larger local Sommerfeld enhancement in the host halo
$S_0 = 1000$ corresponding to  $\alpha = 1/30$.
}
\label{Hostflux}
\vspace{2mm}
\end{figure}

The Sommerfeld-enhanced luminosity functions are steeper, with $s=2.8$
for $S \sim 1/v$, and $s=4.9$ for the $S \sim 1/v^2$ model. In the
smallest mass subhalos resolved in VL2, the internal velocity
dispersions become comparable to the saturation velocity. Even smaller
subhalos will be fully saturated, and the slope of their luminosity
function should be identical to the non-Sommerfeld-enhanced case. We
thus expect a break in the power law right around the saturation
luminosity $L_{\rm sat}$. We estimate $L_{\rm sat}$ by determining the
mean luminosity of all subhalos with $V_{\rm max}/c$ within 10\% of
$v_{\rm min}/c=m_\phi / \mdm=10^{-5}$, and find $L_{\rm sat}=2 \times
10^9 M_\odot^2$ pc$^{-3}$ for $S \sim 1/v$ and $L_{\rm sat}=10^{11}
M_\odot^2$ pc$^{-3}$ for $S \sim 1/v^2$. Since these values are very
close to the VL2 completeness limit, it is difficult to clearly
distinguish this break. For a model with a higher force carrier to DM
particle mass ratio of $m_\phi / \mdm = 5 \times 10^{-5}$, this break
is indeed apparent at the higher saturation luminosity of 
$L_{\rm  sat}=2 \times 10^9 M_\odot^2$ pc$^{-3}$.  These results indicate
that subhalos below the VL2 resolution limit are unlikely to either
strongly enhance or suppress the likelihood of having a significant
spectral feature due to an individual subhalo.

As a concrete example, we consider the electron flux from DM
annihilating in a clump 
at a distance $x_{\rm sub}=1$~kpc.  We take the $S \sim 1 / v$ model with
the normalization $S_0 = S(200 {\rm km/s}) = 600$
and choose the following luminosity of the DM clump
\be
\lb{subh-lum}
L_{\rm sub}
= 10^9 \; M_\odot^2 \; \text{pc}^{-3}.
\ee
This luminosity is equal to the total luminosity in the host halo within approximately
2 kpc from Earth.
$L_{\rm sub}$ is about 4 times smaller than the luminosity resolved in VL2.
Consequently, in the equipartition case $s = 2$, the probability to find a 
clump of DM with $L > L_{\rm sub}$ is around $4 \cdot 4\% = 16\%$.
The source function for the electrons and positrons from DM annihilating 
in a subhalo at position $\vx_{\rm sub}$ is
\be
Q_{\rm sub}(\vx, E) = \kappa L_{\rm sub} \dl^3(\vx - \vx_{\rm sub}) Q(E).
\ee
The flux from a DM subhalo
can be obtained by substituting the source $Q_{\rm sub}$ into the general solution
(\ref{density-soln})
\bea
\nonumber
F_{\rm sub} &=& \frac{c}{4\pi} \frac{1}{b(E)}
\kappa L_{\rm sub}\\
\lb{subhalo_flux}
&&
\cdot\int_E^{\mdm} dE_0\;
 \frac{1}{(4\pi\ld)^{3/2}} e^{-\frac{\vx_{\rm sub}^2}{4\ld}} \;  Q(E_0).
\eea
In our example, we use the DM toy model from section \ref{sec:DM-models}
with the annihilation chain
$2\chi \;\ra\; 2\phi_1 \;\ra\; 4\phi_2 \;\ra\; 4e^+ + 4e^-$.
The corresponding source function $Q_3(E)$ 
was presented in Eq. (\ref{DMmodels}).
This model can be considered as a toy model for a more realistic decay 
through pions or muons instead of $\phi_2$.
If there are four pions (muons) after the decay of $\phi_1$, then there will be only
2 electrons and 2 positrons in the final state and the boost factor in this 
model should be about twice larger than the boost factor in the toy model
with 4 electrons and 4 positrons in the final state.

In Fig. \ref{ClumpsVsHalo}, 
we compare the flux from a subhalo 
with the flux from the host halo
\be
F_{\rm host} = \frac{c}{4\pi} \frac{1}{b(E)}
\int_E^{\mdm} dE_0\; \kappa \bra \rhodm^2 S_0 \ket Q(E_0).
\ee
Since the dependence on the host halo density profile is insignificant,
we use the homogeneous DM distribution with
${\rhodm} = 0.3\:\text{GeV\,cm}^{-3}$ and 
$\bra \sm v \ket_0 = 3 \times 10^{-26}\text{cm}^3\text{s}^{-1}$. 
We have chosen a local boost factor of  $S_0 = 600$ 
and a DM particle mass of ${\mdm} = 2.1$ TeV
 in order to fit the ATIC and PAMELA data 
(see Fig. \ref{1kpcAticPam}).
Thus the flux in the presence of significant contribution from a subhalo
is consistent with the ATIC and PPB-BETS data but may contradict the Fermi data 
if we simultaneously fit to the PAMELA positron ratio.

The flux from the host halo without subhalos  is shown in Fig. \ref{Hostflux},
where we use the same model as above but increase the boost factor to
$S_0 = 900$ in order to fit the PAMELA, Fermi, and HESS points.  For a
smaller DM mass, one can also fit the PAMELA and ATIC data (as was
shown, for instance, in \cite{Malyshev:2009tw}).  Thus either the ATIC
or the Fermi data can be explained by DM annihilation in the host halo
for different parameters in the DM models (but not simultaneously,
since ATIC and Fermi data have significant deviations from each other
at energies between 300 GeV and 700 Gev).
The fits of the flux from the host halo to the Fermi-LAT 
and PAMELA data and the corresponding ranges of parameters
for a more realistic counterpart of our toy model with the annihilation chain
$2\chi \;\ra\; 2\phi_1 \;\ra\; 4\mu^{\pm}$
can be found in \cite{Bergstrom:2009fa}.

\section{Conclusions}
\lb{sec:Concl}

In this paper we have studied the $e^+e^-$ flux from annihilating DM.
We analyzed the dependence of the flux on the choice of DM model, on
the profile of the DM halo, and on the presence of a local subhalo.

The DM models we have considered here are characterized by the mass of
the DM particle $\mdm$ and by the number of step $k$ in the
annihilation-decay process from DM to $e^+e^-$. In particular, we
studied a toy model where the DM particles annihilate into two scalars
that decay by a chain of (k - 1) two-body decays into electrons and
positrons.  In this model, $\log \frac{\mdm}{E_*} \approx \frac{k
  - 1}{2}$ and, as expected, $F \sim E^{-2}$ for $E \ll E_*$.  The
behavior of the flux for $E \sim E_*$ is model dependent.  The
spectrum has a sharper cutoff in models with fewer decay steps.

The dependence on the shape of the DM host halo is very mild, provided
that the DM distribution is sufficiently smooth.  For the typical DM
halo profiles, such as NFW, Einasto, and Isothermal, $\dl n \lesssim
10\%$.

If a small number of DM subhalos contribute significantly to the DM
annihilation, then the corresponding source function may have large
variations and the index of the propagated $e^+e^-$ flux will change
as well.  In general the index grows, i.e., the flux from a subhalo is
harder than the flux from the host halo.  Using the {\VLII} simulation,
we argue that, on average, we expect one $M>10^5 M_\odot$ subhalo
within 3 kpc from the Earth, and the flux from such a subhalo will
leave a significant imprint on the electron spectrum above $\sim 100$
GeV, if its distance is less than $\sim 1$ kpc.  Thus, based on the
\VLII simulation, there is at least a $\sim 4\%$ chance to
observe the flux of electrons and positrons from a local DM subhalo.
Extrapolating below the VL2 resolution limit, we estimate that this
probability may grow to $\sim$15\%.

In the presence of DM subhalos, there may be some features in the
$e^+e^-$ flux spectrum at high energies, but at low energies we should
still expect the universal index $n \approx -2$ of the DM annihilation flux. 
Future observations will help to distinguish between the following possibilities:
\bi
\item
The additional flux is dominated by a local DM clump at all energies.
Then the index of the flux is not universal and, as a rule,
$n > -2$, i.e. the flux is harder than the flux from the host halo.
However, as was pointed out in \cite{Malyshev:2009tw}, 
we need a flux with an index $n = -2.2\pm0.2$
in order to fit both ATIC and PAMELA data.
This problem becomes even worse if one tries to 
fit simultaneously the PAMELA points and
rather soft Fermi/LAT spectrum.
Therefore this possibility contradicts current data.
\item
The additional flux is dominated by a local DM clump 
at high energies, 100 - 1000 GeV,
and by the host halo at low energies,
10 - 100 GeV.
We present an example with a DM clump at 1 kpc from the Earth
consistent with the ATIC and PAMELA data.
However, this flux may be inconsistent with the Fermi/LAT and 
the HESS data.
The flux from annihilating DM has an index
$n \approx -2$ at energies $E < 100$ GeV and $n > -2$ at energies 
100 GeV $< E <$ 300 GeV. 
\item
The additional flux is dominated by the host halo at all energies.
For different DM models
this flux can fit either PAMELA and ATIC or PAMELA and Fermi data but not both.
The property that the positron ratio is fitted in both cases
is a consequence of a general fact
that,
independently of DM model and the shape of the DM halo profile,
we expect an index $n \approx -2$ for energies much smaller than the cutoff scale $E \ll E_*$.
\ei

Our main conclusion is that at lower energies $E \ll \mdm$, i.e. in
the PAMELA range 10 - 100 GeV for $\mdm \gtrsim 1$ TeV, one should
expect the flux from the DM to have a universal index $n \approx -2$.
At higher energies, i.e., in the ATIC - Fermi range 100 - 1000 GeV,
the behavior of the flux is model dependent.  In the presence of a
significant nearby clump of DM one should expect an ATIC-like bump in
the spectrum whereas in the absence of large nearby clumps the flux
should look smooth, similar to the Fermi/LAT and HESS data.

\bigskip
\bigskip

\noindent
{\large \bf Acknowledgments.}
\medskip

\noindent
The authors are thankful to J. Bovy, I. Cholis, J. Diemand, L.
Goodenough, J. Roberts, and N. Weiner for valuable
discussions. This work is supported in part by the Russian Foundation
of Basic Research under Grant No. RFBR 09-02-00253 (DM), by NSF Grant Nos.
PHY-0245068 (DM) and PHY-0758032 (DM). MK acknowledges support from
the William L. Loughlin Fellowship at the Institute for Advanced
Study.


\bibliography{DMpapers}         
\bibliographystyle{utphys}   

\end{document}